\documentclass[conference]{IEEEtran}
\usepackage{boldline}
\usepackage{xcolor}
\newcolumntype{L}[1]{>{\raggedright\let\newline\\\arraybackslash\hspace{0pt}}m{#1}}
\newcolumntype{$}{>{\global\let\currentrowstyle\relax}}
\newcolumntype{^}{>{\currentrowstyle}}
\newcommand{\rowstyle}[1]{\gdef\currentrowstyle{#1}%
  #1\ignorespaces
}

\usepackage{comment}

\usepackage{cite}

\ifCLASSINFOpdf
   \usepackage[pdftex]{graphicx}
\else
   \usepackage[dvips]{graphicx}
\fi

\usepackage{amsmath}

\usepackage{array}

\ifCLASSOPTIONcompsoc
  \usepackage[caption=false,font=normalsize,labelfont=sf,textfont=sf]{subfig}
\else
  \usepackage[caption=false,font=footnotesize]{subfig}
\fi
\usepackage[hyphens]{url}

\usepackage{bbm}
\usepackage{fixltx2e}
\usepackage{algorithm}
\usepackage[noend]{algpseudocode}
\usepackage[outercaption]{sidecap}

\usepackage[section]{placeins}

\MakeRobust{\Call}

\makeatletter
\def\BState{\State\hskip-\ALG@thistlm}
\makeatother

\usepackage{mathtools}
\DeclarePairedDelimiter{\ceil}{\lceil}{\rceil}

\algnewcommand\Or{\textbf{ or }}

\makeatletter
\renewcommand\fs@ruled{%
  \def\@fs@cfont{\rmfamily}%
  \let\@fs@capt\floatc@plain%
  \def\@fs@pre{\hrule height.8pt depth0pt \kern2pt}
  \def\@fs@post{}
  \def\@fs@mid{\kern2pt\hrule\kern2pt}
  \let\@fs@iftopcapt\iffalse}
\makeatother

\sidecaptionvpos{figure}{c}

\begin{document}
\title{When the cookie meets the blockchain: \\
Privacy risks of web payments via cryptocurrencies 
}

\author{\IEEEauthorblockN{Steven Goldfeder\IEEEauthorrefmark{1}, Harry Kalodner\IEEEauthorrefmark{1}, Dillon Reisman\IEEEauthorrefmark{2}, Arvind Narayanan\IEEEauthorrefmark{1}}
\IEEEauthorblockA{Princeton University}
\IEEEauthorblockA{\IEEEauthorrefmark{1}\{stevenag, kalodner, arvindn\}@cs.princeton.edu}
\IEEEauthorblockA{\IEEEauthorrefmark{2}dillon@lonlon.io}}


%

\maketitle
\thispagestyle{plain}
\pagestyle{plain}

\begin{abstract}
We show how third-party web trackers can deanonymize users of cryptocurrencies. We present two distinct but complementary attacks. On most shopping websites, third party trackers receive information about user purchases for purposes of advertising and analytics. We show that, if the user pays using a cryptocurrency, trackers typically possess enough information about the purchase to uniquely identify the transaction on the blockchain, link it to the user's cookie, and further to the user's real identity. Our second attack shows that if the tracker is able to link two purchases of the same user to the blockchain in this manner, it can identify the user's entire cluster of addresses and transactions on the blockchain, even if the user employs blockchain anonymity techniques such as CoinJoin. The attacks are passive and hence can be retroactively applied to past purchases. We discuss several mitigations, but none are perfect.

\end{abstract}




\section{Introduction}
\label{sec:introduction}
Eight years after Bitcoin's introduction, the ability to pay online using cryptocurrencies is common: prominent merchants such as Microsoft, Newegg, and Overstock support it. Cryptocurrency users tend to value financial privacy, and it is a major reason for choosing to pay with Bitcoin \cite{krombholz2016other}. Yet, websites including shopping sites are known to be rife with third-party tracking \cite{englehardt2016online}. In this paper, we study the impact of online tracking on the privacy of Bitcoin users.

\begin{figure*}
\begin{centering}

  \includegraphics[height=6cm]{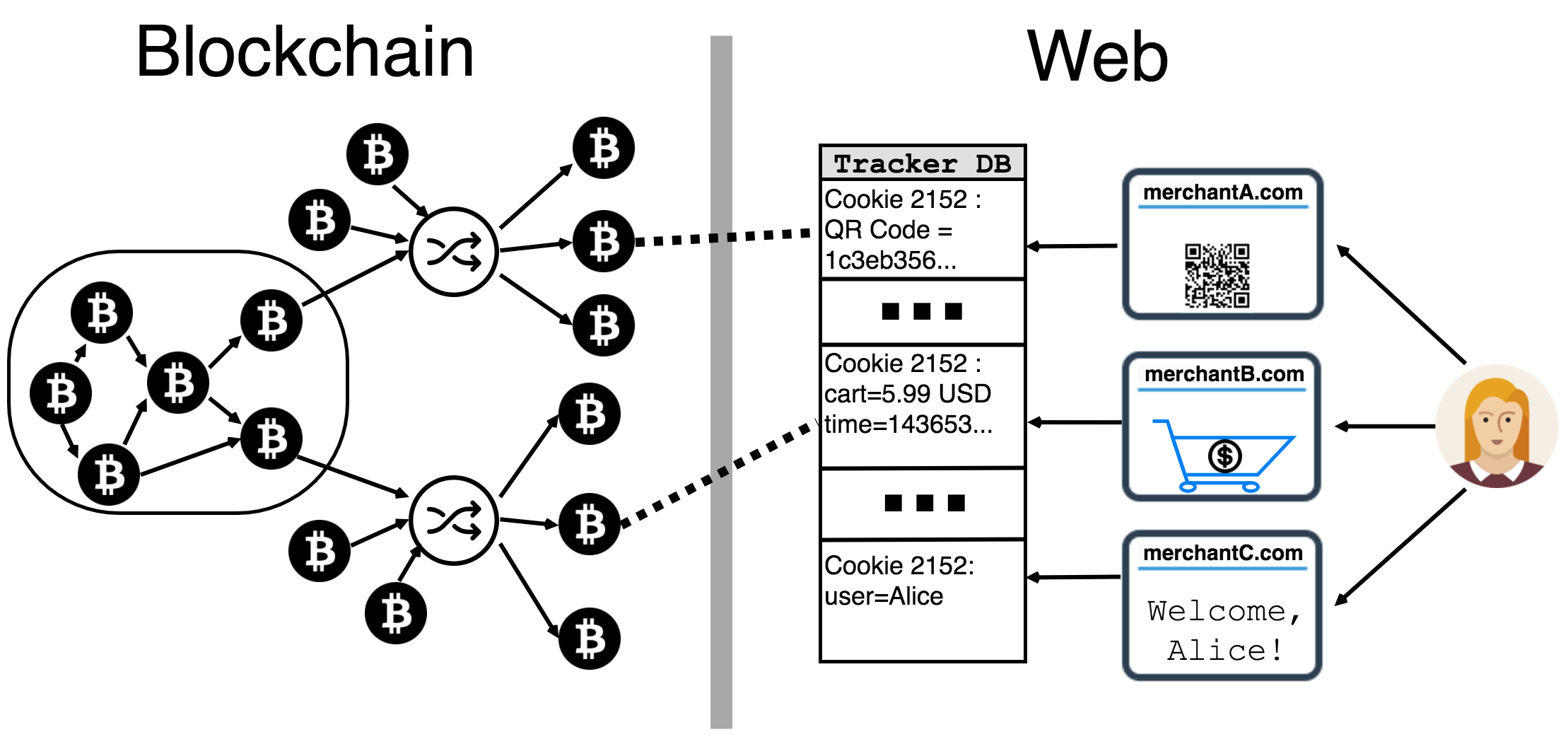}
  \caption{{\bf An illustration of the full scope of our attack.}} Consider three websites that happen to have the same embedded tracker. Alice makes purchases and pays with Bitcoin on the first two sites, and logs in on the third. Merchant A leaks a QR code of the transaction's Bitcoin address to the tracker, merchant B leaks a purchase amount, and merchant C leaks Alice's PII. Such leaks are commonplace today, and usually intentional (Section \ref{sec:webmeasurement}). The tracker links these three purchases based on Alice's browser cookie.  Further, the tracker obtains enough information to uniquely (or near-uniquely) identify coins on the Bitcoin blockchain that correspond to the two purchases. However, Alice took the precaution of putting her bitcoins through CoinJoin before making purchases. Thus, either transaction individually could not have been traced back to Alice's wallet, but there is only one wallet that participated in both CoinJoins, and is hence revealed to be Alice's.
    \label{fig:attack-diagram}
  \end{centering}
\end{figure*}

First, we show that online trackers are able to see sensitive details of payment flows, such as the identities and prices of items added to shopping carts. Crucially, in many cases they receive sufficient information about a purchase to link it uniquely to a transaction on the Bitcoin blockchain.\footnote{Throughout we study Bitcoin since it has the most support for online payments, but our findings apply to many other cryptocurrencies.} This core linkage can be expanded in both directions: based on tracking cookies, the transaction can be linked to the user's activities across the web. And based on well-known Bitcoin address clustering techniques \cite{meiklejohn2013fistful,androulaki2013evaluating}, it can be linked to their other Bitcoin transactions.

This basic attack can be made worse in several ways. We find that many merchant sites send even more information to trackers, such as the transaction-specific Bitcoin address. This acts as a high-entropy identifier and makes linking to the blockchain trivial. We also show that many merchants additionally leak users' PII (name, email address, etc.) to trackers, allowing trackers to link not only users' web profiles but also blockchain transactions to their identities. Finally, malicious trackers may use JavaScript to extract Bitcoin addresses or PII from web pages even if it is not leaked to them by default. We show that this is possible on the vast majority of merchant sites.

Of course, Bitcoin does not guarantee unlinkability of transactions. But while linking of a user's Bitcoin addresses {\em with each other} is well known \cite{meiklejohn2013fistful,reid2013analysis,ron2013quantitative,androulaki2013evaluating}, our attack shows how to link addresses to external information, including identity. 

The main defense against linkage attacks is mixing \cite{Bonneau2014,ruffing2014coinshuffle}. The best known mixing technique is CoinJoin, in which users send coins to each other in a way that hides the link between their old and new coins. Our second main contribution is showing the effectiveness of the {\em cluster intersection attack}, a previously known attack against mixing. Specifically, we show that a small amount of additional information, namely that two (or more) transactions were made by the same entity, is sufficient to undo the effect of mixing (see Figure \ref{fig:attack-diagram}). While such auxiliary information is available to many potential entities --- merchants, other counterparties such as websites that accept donations, intermediaries such as payment processors, and potentially network eavesdroppers --- web trackers are in the ideal position to carry out this attack.

Based on the above two attacks, we present the following findings. We present a taxonomy of information leaks to trackers on e-commerce websites. We focus on leaks that allow linking a payment flow to a blockchain transaction.  We compiled a list of 130 online merchants that accept Bitcoin, and analyzed their websites by extending the functionality of the open-source OpenWPM web privacy measurement tool \cite{englehardt2016online}. We find that at least 53/130 of merchants leak payment information to a total of at least 40 third parties, most frequently from shopping cart pages. The vast majority of these represent {\em intentional} sharing of purchase data with third parties for advertising and analytics purposes. In addition, we find that many merchant websites have far more serious (and likely unintentional) information leaks that directly reveal the exact transaction on the blockchain to dozens of trackers. 

Turning to the Bitcoin blockchain, we use empirical measurement to estimate the uniqueness of transactions as a function of the adversary's uncertainty about
the transaction's timestamp and value (Section \ref{sec:blockchain}). We find that unique linkage is possible in over 60\% of cases for realistic values of these parameters, and that in the vast majority of cases, the anonymity set size is 5 or less. The attack degrades gracefully as the adversary's uncertainty increases. Note that in the case of the unintentional leaks mentioned above, there is no uncertainty, and unique linkage is always possible.

Next, we evaluate the efficacy of the cluster intersection attack  against CoinJoin (Section \ref{sec:cluster}; in Section \ref{sec:mitigation} we discuss the applicability to other types of mixing). By identifying a corpus of 78,697 CoinJoin transactions on the Bitcoin blockchain over a two-year period, we present realistic simulations of a victim who mixes coins from her wallet and then makes payment transactions that are observed by the adversary. 
For example, if the victim employs 3 rounds of CoinJoin and the adversary observes two of the victim's payments, he can link them back to her wallet (despite mixing) with 98\% accuracy. Multiple rounds of mixing increase privacy, but those gains are quickly stripped away if the adversary observes more than 2 payments.


Finally, we evaluate our attack end to end (Section \ref{sec:end-to-end}). We made 21 purchases on 20 merchant websites. For 11 of these purchases, we used freshly mixed coins to attempt to deter linkage. There were 25 pairs of purchases made with mixed coins for which there was at least one tracker that received leaked data about both purchases. We find that in 20 of these 25 cases, the tracker can identify the user's wallet despite the use of mixing.

Our attack highlights the dangers of pervasive web tracking: Bitcoin is often used for sensitive activities, making the compromise of Bitcoin privacy a far more serious threat than targeted advertising. In Section \ref{sec:mitigation} we discuss mitigations that merchants can deploy. None is a complete solution, given the fundamental tension between privacy and the analytics needs of modern e-commerce. Indeed, most of the privacy-breaching data flows we identify are intentional and not accidental (Section \ref{sec:webmeasurement}).

The main self-defense available to users today is to use tracking-protection tools such as Ghostery or uBlock Origin, but we note several limitations. First, since our attack is passive, trackers have {\em already} accumulated data in their logs that enable them to retrospectively carry out the attack. Second, tracking protection tools aren't perfect and contain both false positives (resulting in broken functionality) and false negatives (resulting in missed trackers). In Section \ref{sec:webmeasurement} we show that even with tracking protection enabled, 25 merchants still leak sensitive information to third parties. Third, merchants, payment processors, and even network eavesdroppers are potential adversaries for some of the attacks we describe, and tracking protection does not help against these adversaries. Finally, in Section \ref{sec:mitigation} we also discuss how our techniques can aid law enforcement investigations.












\section{Background and Related Work}
\label{sec:background}
Our work brings together two previously unrelated areas of privacy research: web tracking and anonymity of cryptocurrencies. We describe each in turn.

{\bf Online tracking.} Since the web's inception, the number of third parties that track and record user activity has exploded. \cite{lerner2016internet,mayer2012third,roesner2012detecting,budak2016}. In this paper we use the terms third party and tracker interchangeably. Some trackers have a substantial view of users' activities across the web: Google, for instance, has a tracking presence on roughly 80\% of sites \cite{libert2015exposing}. Tracking methods have also become  more sophisticated over time \cite{soltani2010flash, englehardt2015cookies,acar2014web,eckersley2010unique,laperdrix2016beauty}. The effectiveness of tracker-blocking tools has been studied by various authors \cite{yu2016tracking,merzdovnik2017block,gervais2016quantifying}.

Some trackers like Google and Facebook are known to tie their tracking profiles to identities directly disclosed by users, but most trackers have no direct relationship with users. However, even such trackers acquire PII, often accidentally. Various studies starting in 2009 have shown that the leakage of PII from first parties to third parties is rampant \cite{krishnamurthy2011privacy,krishnamurthy2009leakage}, and the problem remains severe today. 

Most trackers are legitimate businesses, but are known to use intrusive means to track users. These include misuse of HTML5 APIs for fingerprinting, such as Canvas, Audio Context, and Battery Status \cite{englehardt2016online}; cross-device tracking \cite{xdt}; workarounds to browser privacy features \cite{angwin2012google}, and sniffing data from unsubmitted forms \cite{navistone}. Many trackers have poor security on their servers and are a target for compromise for malvertising and other purposes \cite{nikiforakis2012you,sood2011malvertising}.

The problem of trackers observing shopping and payment flows is unlikely to go away. Consider retargeting, which is the ability to serve ads to users for items they are known to have shown an interest in purchasing. It is one of the most valuable forms of advertising \cite{retargeting}. The farther into a payment flow a tracker can observe a user (cart page, checkout page, etc.) the greater the interest signaled. Another major benefit is conversion tracking of ad campaigns. Having trackers on the payment flows is needed to help analyze whether a user who was served an ad actually follows through with a purchase. Other applications include fraud/abuse detection and consumer insights.

{\bf Cryptocurrencies.} In Bitcoin-like cryptocurrencies, users pay by broadcasting transactions to a peer-to-peer network. Transactions are signed statements authorizing transfers from one address to another. Addresses are public keys that act as pseudonymous ``account" identifiers. Transactions are recorded in an immutable, global ledger called the blockchain \cite{Nakamoto2008, BMCNKF15}.

{\bf Address clustering and mixing.} It is trivial to generate new Bitcoin addresses, and most wallet software takes advantage of this feature to improve user privacy. In the normal course of operation, users end up with coins split between numerous addresses, and it may not be obvious which addresses belong to the same user (or entity). However, there are well known and well understood attacks to infer links between such addresses \cite{meiklejohn2013fistful,reid2013analysis,ron2013quantitative,androulaki2013evaluating}. These techniques have been improved upon and implemented by companies such as Chainalysis and made available via easily accessible APIs. Address clustering is not perfect, but it is a powerful attack, and wallet addresses must be considered clusterable unless additional privacy-protection techniques are employed to break the link between those addresses.

Many such privacy-protection techniques are known \cite{anonsurvey}; the ones readily deployable on existing Bitcoin-like cryptocurrencies are all variants of the idea of {\em mixing}. The best known and most used technique is known as CoinJoin \cite{CoinJoin,ruffing2014coinshuffle}, in which different users coordinate in order to jointly create a transaction that spends a coin of equal value from each of them, and from which each of them receives a coin of the same value. The order of outputs is randomly permuted so that the mapping between inputs and outputs cannot be deduced from the public blockchain. Services such as JoinMarket provide the ability for users to coordinate to mix their coins \cite{moser2016join}.

CoinJoin improves unlinkability by breaking the multi-input heuristic, one of the main heuristics used in address clustering. However, the susceptibility of CoinJoin (and other mixing techniques) to clustering has not yet been rigorously studied. It is known that CoinJoin transactions are at least {\em detectable} as such, since they involve many inputs and outputs with the same value, a highly unlikely pattern in a regular payment transaction. In other words, CoinJoin improves anonymity but does not provide unobservability \cite{moser2016join}.

Intersection attacks date back to the communications anonymity literature and are well known. Their applicability to cryptocurrency mixing is also generally understood. At least two papers mention it explicitly \cite{bissias2014sybil,heilman2016blindly}, but they focus on mix participants and other intermediaries as adversaries. A 2015 blog post also mentions the attack \cite{ordano2015need}. What's new in our work is the idea that auxiliary information to link different mixed coins is readily available to web adversaries (as opposed to behavioral patterns in earlier work, which is a much less reliable linkage mechanism). Further, we are able to empirically evaluate the attack using recently proposed techniques for identifying CoinJoin transactions on the blockchain \cite{moser2016join} (Section \ref{sec:cluster}).

{\bf Other research on cryptocurrency privacy and forensics.} Gervais et al. present an intriguing attack on e-commerce purchases using cryptocurrencies: since prices are denominated in local currencies, and are usually close to integer multiples of the unit of currency, blockchain transaction amounts could reveal the currency and hence the location of the purchase \cite{gervais2015quantifying}. Our work is complementary; their attack is stronger than ours in that the adversary can be anyone examining the blockchain, whereas our attack is stronger in the sense that much more information is leaked, and not just the location.

Another major route to compromise of cryptocurrency privacy, orthogonal to ours, is the linkage of transactions to the sender's IP address. An adversary who is well connected to the Bitcoin peer-to-peer network might be able to do so~\cite{DBLP:conf/fc/KoshyKM14, biryukov2014deanonymisation}; even users who connect to the Bitcoin network over Tor are potentially vulnerable  \cite{biryukov2015bitcoin}. In response to these attacks, Bitcoin Core changed the protocol for how transactions are disseminated across the network in 2015. However, recent work showed weaknesses in the updated protocol \cite{fanti2017anonymity,juhasz2016bayesian,neudeckercould}. A re-designed P2P networking protocol with strong anonymity guarantees has been proposed \cite{venkatakrishnan2017dandelion}, but not yet adopted by any cryptocurrency.

In concurrent work, Portnoff et al. explore a technique similar to our transaction linkage attack \cite{portnoff17backpage}. In their work, linkage is a forensic technique to help identify entities behind illegal activities (sex trafficking). It is enabled by a specific feature of a specific website, backpage.com: classified ads paid for by users are posted on the website along with an accurate timestamp. This allows anyone (e.g., researchers, NGOs, law enforcement) to link an ad to the transaction on the bitcoin blockchain that represents the payment for the ad. In our work, linkage can be carried out only by specific entities, such as trackers, but we extend the linkage via cookies, PII, and blockchain analysis, none of which are applicable to the setting of Portnoff et al. Of course, their work can be viewed as a demonstration of a privacy breach affecting Backpage users, including the majority not engaging in illegal activities; similarly, our attack can be turned around into a forensic technique (Section \ref{sec:mitigation}).
 


\section{Threat model and attacks}
\label{sec:threatmodel}
{\bf Merchant, payment processor, and trackers.} A typical cryptocurrency-based e-commerce flow consists of a user, a merchant, a payment processor, and one or more trackers. The merchant is the website where the user is shopping. Most merchants make use of payment processors such as BitPay and Coinbase to handle the details of processing cryptocurrencies. When the user pays with Bitcoin or another cryptocurrency, the transaction is received by the payment processor, who then usually credits the merchant's account with an equivalent amount of dollars or other local currency. Trackers are ``third parties'' on web pages, often invisible, that track users' actions for purposes of advertising, analytics and so on (Section \ref{sec:background}). Doubleclick, Google Analytics, and Facebook are common examples. Merchants, payment processors, and trackers are all potential adversaries in our attack, although we are most interested in the latter.

\begin{figure*}
\begin{centering}
  \includegraphics[height=2.8cm, trim={0 2cm 0 1cm},clip]{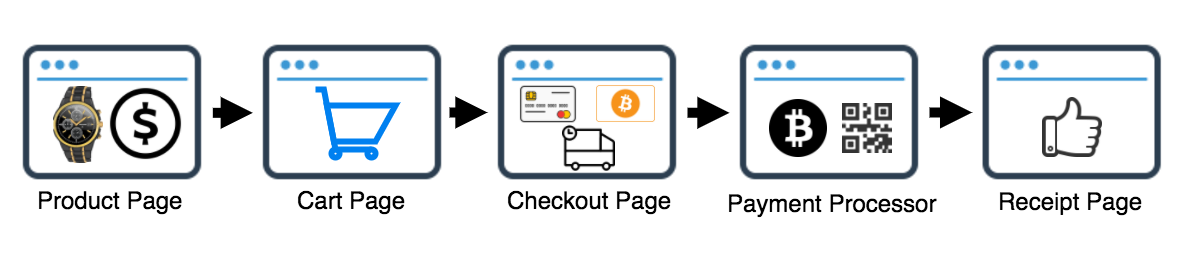}
  \caption{{\bf An illustration of a typical payment flow on a merchant site.} Each step of this flow presents opportunities for leaking transaction-relevant information to embedded third-party trackers.}
    \label{fig:payment-flow}
  \end{centering}

\end{figure*}

{\bf Information flows to third parties.} Users take actions on shopping sites such as logging in, viewing items, adding items to their cart, checking out, and making a payment. See Figure \ref{fig:payment-flow} for an illustration of a typical payment flow on a merchant site. The more of these actions a third party learns about, the more feasible the attack. The types of information useful to the third party are:
\begin{itemize}
\item Payment timestamp: the third party learns the approximate payment time simply by virtue of being embedded on merchant website, especially on pages constituting the checkout process. Checkout pages usually require the user to complete payment (i.e., broadcast the cryptocurrency transaction) within a short time window, typically 15 minutes. Trackers embedded on payment receipt pages are in an even better position, as they learn the payment time to within a few seconds. Note that assuming the user included a reasonable transaction fee, payment processors consider payments received as soon as the transaction is broadcast to the peer-to-peer network and received by the payment processor's node. This involves a latency of only a few seconds. The transaction may not be {\em confirmed} until it is incorporated into the blockchain, which may take tens of minutes depending on the degree of confirmation that the payment processor requires. The transaction confirmation time is largely irrelevant to our attack.
\item Payment address: the payment address is the destination to send coins. Recipients (payment processors) will typically generate a fresh address specific to the transaction --- new Bitcoin addresses are trivial to generate. Although there is no business reason for trackers to receive the payment address, we find that this does happen often (Section \ref{sec:webmeasurement}). Since payment addresses are unique, at least within the time scale of interest to us, a leak of the payment address trivially allows the tracker to link the web user to the blockchain transaction.
\item Price: Depending on the merchant website, trackers may be able to see the prices of items viewed by users, items added to the cart, or even the final price after shipping and taxes have been calculated. Note that these prices are almost always denominated in USD, EUR, or another fiat currency, even on websites that only accept cryptocurrencies as payment, due to the extreme volatility of cryptocurrency exchange rates. However, once the user checks out, the amount is calculated in BTC (or another cryptocurrency) based on the exchange rate at that instant. In some cases, this BTC-denominated price is also revealed to the tracker, which is more useful for linking than the price before applying the conversion.
\item Personally Identifiable Information (PII): By PII we mean any information related to the user's real identity or account on the merchant website, such as name, email address, username, and shipping address. Trackers' access to PII exacerbates the attack. 
In this paper we analyze leaks of PII from merchant websites to trackers, but we emphasize that since trackers are widely present on the web, the link to PII can be acquired on any website whatsoever. Leaks of PII to trackers are known to be rampant across the web (Section \ref{sec:background}).
\end{itemize}

In our measurements in Section \ref{sec:webmeasurement}, we focus on passive attacks where trackers obtain this information in the normal course of operation. Except for (some) PII leaks, most other information flows to trackers are intentional: trackers use this information for advertising and analytics purposes. However, we note that in many cases, tracking scripts are in a position to carry out an active attack and extract all of the above information from web pages even if they don't obtain it passively. This is because third-party scripts are typically embedded without any isolation, in a way that gives them full access to the content on the page. Sandboxing techniques such as iframes are readily available, but only infrequently employed since they interfere with some of the functionality provided by trackers.

{\bf Attack 1: single transaction linkage.} In this attack, the adversary (tracker) seeks to link a web user (as identified by the user's cookie or PII) to a transaction on the cryptocurrency blockchain. The merchant and payment processor are not interesting adversaries for this attack, because it is unsurprising that they can carry out this linkage (but see Section \ref{sec:other-adversaries}). We assume that the user is aware of this possibility, and potentially takes necessary precautions, such as mixing to unlink the transaction on the blockchain from her {\em other} blockchain transactions and addresses. Attack 2 seeks to overcome such defenses. But the tracker's ability to link to even a single transaction is a privacy breach, because the user has no business relationship with the tracker and many users are in fact unaware of the existence of trackers (or at least their prevalence and sophistication). It is also worrisome because trackers compile profiles of users' activities across the web.

If the tracker has access to the receiving address, it trivially enables linkage, as noted above. The more interesting case is when the tracker knows the approximate price and time. Then the tracker's task is to search the logs of transactions that were broadcast to the peer-to-peer network to identify those that fall within the window of uncertainty both in terms of transaction value and time. To quantify the tracker's success, then, we must model the uncertainty in the tracker's knowledge of price and time.

\begin{itemize}
\item Price uncertainty: The tracker's uncertainty around price arises primarily from shipping. If the tracker knows the adversary's location (either based on a leak of PII or based on IP address), this uncertainty can be minimized, although there might still be a small number of possible values of the shipping fee based on the shipping speed selected by the user.
\item Exchange rate uncertainty: The second source of uncertainty is the exchange rate: the tracker usually sees prices denominated in USD (or another fiat currency) and not in BTC. Most payment processors use exchange rates based on trading data publicly released by an exchange, which means the tracker can always reconstruct the exchange rate at any given point in the past. However, since trades happen several times per second, the exchange rate varies rapidly and hence some uncertainty will still remain. 
\item Payment time uncertainty: this arises because of the gap between the user checking out, the user's wallet broadcasting the transaction, and that broadcast being recorded by the adversary or another node. The adversary may run his own peer-to-peer Bitcoin node, or may simply obtain the transaction broadcast timestamp from publicly available sources such as blockchain.info. If the tracker is present on the transaction receipt page, then the latency is minimized, and is of the order of the network propagation delay, i.e., a few seconds.
\end{itemize}

{\bf Attack 2: Cluster intersection.} This is a complementary attack where the adversary aims to identify the cluster of addresses in the victim's Bitcoin wallet. Recall from Section \ref{sec:background} that wallets can (and do) easily create numerous addresses, but in the normal course of operation these addresses can still be linked together via various heuristics. Mixing techniques such as CoinJoin are thought to protect against such linkage, although this has not yet been studied rigorously. We assume that the victim uses a desktop (local) wallet rather than an online wallet provider. Privacy-conscious users tend not to use online wallets, since that would allow the wallet provider to trivially track all of the user's activities. We also assume that the user employs effective communications anonymity techniques to mask the IP addresses of their wallet, as that is a well-known way for anonymity to be compromised (Section \ref{sec:background}).
 
In our attack, the victim interacts with the adversary multiple times. The adversary could be a merchant, payment processor, or (especially) a tracker who only indirectly observes the victim. Knowing that the adversary might learn one of his addresses, the victim employs mixing to prevent the adversary from learning the rest of his addresses and transactions. He doesn't spend coins directly from his wallet, but only after first mixing them. In Figure \ref{fig:attack-diagram}, after the victim has shopped on {\em merchantA.com} the adversary is unable to determine which of the three wallet clusters belongs to the victim. But after a second interaction with the same victim on {\em merchantB.com}, the adversary simply finds the intersection of the two sets of clusters, which leads him to a unique cluster.


Web trackers passively observe users' web purchases and are able to link them together, via cookies or device fingerprinting, even if the merchant and payment processor are different in every case. Thus, this attack is complementary to Attack 1, and would take as input two blockchain transactions identified via Attack 1. Note that even if Attack 1 is imperfect, and returns a set of transactions instead of a single one, Attack 2 will still succeed. The intersection size rapidly decreases as a function of the number of observations, and even if two observations aren't sufficient to uniquely identify the wallet, it is likely that a small number of additional observations will suffice. We quantify this in Section \ref{sec:cluster}.

\section{Web measurement: Leaks of sensitive data}
\label{sec:webmeasurement}
In this section we analyze leaks of sensitive data on merchant sites. In sections \ref{sec:blockchain} and  \ref{sec:end-to-end} we examine how trackers can actually use this data to identify transactions on the blockchain. We also show in this section how trackers can connect this information to users' identities.

\subsection{Method}
To identify leaks of sensitive data, we performed a web crawl of popular merchants that accept Bitcoin. To create a list of merchant sites,  we began by combining popular community-maintained lists of merchants \cite{spendabit,redditbitcoinlist}, which gave us 1438 sites.
We then pruned the list to those domains that were found in the Alexa top 1 million websites, which left 283 sites. As we crawled the sites, we discovered that about half of the merchants no longer accepted Bitcoin. This left 130 merchants in our crawl that accepted Bitcoin at the time of our measurements, and we focus on these 130 sites. These merchants were geographically distributed over 21 countries, with 64 based in the United States and 20 based in the United Kingdom.

Typical merchant payment flows allow us to complete most of the steps  --- viewing products, adding them to the cart, initiating checkout, and receiving a payment address and price --- before actually having to make a payment. This allowed us to collect data on almost the entire payment flow on a large number of websites. However, to analyze payment receipt pages, we need to actually make purchases. Thus we analyze transaction receipt pages on a smaller scale based on actual purchases. We made purchases from 20 distinct merchants in total. 

To collect data on web tracking we used a modified version of the open-source web privacy measurement tool OpenWPM \cite{englehardt2016online}. Using the tool we collected all HTTP(S) requests and responses. We also manually marked any PII and payment-related information that we encountered on the pages we visited; we added functionality to the tool to automatically record this information when marked. 

Throughout our measurements, we are interested in the privacy risk both for a regular user and for a user who employs tracking-protection tools. Most such tools (e.g., Adblock Plus, uBlock Origin) use standard, community-maintained {\em filter lists}: EasyList and EasyPrivacy. To measure the privacy-risk for users of tracking protection lists, we simply re-run our analysis after deleting those third-party URLs in our crawl databases that appear in the lists.

\subsection{Findings}

Based on our measurements of 130 Bitcoin-accepting merchants, we found numerous third parties that receive transaction-relevant information by virtue of their business relationship with the merchant in the normal course of a transaction. We define transaction-relevant information as any information that could help identify the transaction on the blockchain. These potential adversaries could retroactively perform the transaction-linkage attack using data already present in their HTTP logs or databases. We measure information received through either unintentional leakage, via the referer field of an HTTP GET or POST request \cite{krishnamurthy2009leakage}, or through intentional information sharing via an HTTP POST action or a GET URL parameter.


\subsubsection{Third parties that receive Bitcoin address or BTC-denominated price}
Almost all merchants use third-party payment processors; this helps them avoid the security, volatility, and legal risks of receiving and holding bitcoins. Table~\ref{fig:payment-processor-freq} lists the prevalence of different payment processors in our crawl of bitcoin-accepting merchants. The payment processor either sits in an \texttt{iframe} on the checkout page, or on a separate page in the payment flow. The frame or page will display the exact Bitcoin amount the user should send, to an address controlled by the payment processor. 

\begin{table}[]
\centering
\begin{tabular}{|l|l|}
\hline
BitPay      & 70 \\ \hline
Coinbase        & 24   \\ \hline
Coinpayments & 3  \\ \hline
Stripe      & 2 \\ \hline
Other  & 31  \\ \hlineB{4}
Total & 130 \\ \hline
\end{tabular}
\caption{Prevalence of payment processor in 130-site crawl}
\label{fig:payment-processor-freq}
\end{table}

\textbf{We found that 17 of the 130 Bitcoin-accepting merchant websites send the receiving Bitcoin address or BTC-denominated price to a third party} (Table~\ref{fig:btc-tx-info-merchant-leak}). With this information, linking the payment to the blockchain is trivial. The leaks were found on less-popular payment processors and websites that implement their own Bitcoin payment processing. 

We can also break it down by third parties instead of merchants: see Table~\ref{fig:btc-tx-info-third-party-leak}. In both tables, we also present the corresponding measurements for users of tracking-protection tools (via the EasyList and EasyPrivacy\footnote{https://easylist.to/} filter lists). Those results are presented in the ``with protection'' columns. Table \ref{fig:third-parties-despite-tp} in Appendix \ref{app:tables} lists the third parties that receive transaction-relevant information despite the use of tracking protection.

\begin{table}[]
\centering
\begin{tabular}{$L{3.5cm}^L{2cm}^L{2cm}}
\hline
\rowstyle{\bfseries}
Info type & w/o protection & w/ protection \\ \hlineB{6}
Non-BTC-denominated price, incl. shipping & 24 & 12 \\ \hline
Non-BTC-denominated price, pre-shipping & 23 & 5 \\ \hline
Non-BTC-denominated price, either & 43 & 16 \\ \hline
Bitcoin address      & 12 & 12 \\ \hline
Bitcoin price        & 11  & 9 \\ \hline
Bitcoin address or price     & 17  & 15 \\ \hline
Add-to-cart events & 28 & 2 \\ \hlineB{4}
\rowstyle{\bfseries} Total merchants sharing info & 53 & 25 \\ \hline 
\end{tabular}
\caption{Number of merchant sites sending transaction-relevant information to third parties, with or without tracking protection}
\label{fig:btc-tx-info-merchant-leak}
\end{table}

\begin{table}[]
\centering
\begin{tabular}{$L{3.5cm}^L{2cm}^L{2cm}}
\hline
\rowstyle{\bfseries} Info type & w/o protection & w/ protection \\ \hlineB{6}
Non-BTC-denominated price, incl. shipping & 29 & 11 \\ \hline
Non-BTC-denominated price, pre-shipping & 18 & 3 \\ \hline
Non-BTC-denominated price, either & 38 & 13 \\ \hline
Bitcoin address      & 5 & 4 \\ \hline
Bitcoin price        & 4 & 2  \\ \hline
Bitcoin address or price & 9  & 6 \\ \hline 
Add-to-cart events & 9 & 2 \\ \hlineB{4}
\rowstyle{\bfseries} Total third parties receiving info & 40 & 13 \\ \hline
\end{tabular}
\caption{Number of third parties receiving transaction-relevant information, with or without tracking protection}
\label{fig:btc-tx-info-third-party-leak}
\end{table}


On 11 out of the 12 websites that leak the Bitcoin address, the leaks were to third-party services that \emph{render QR codes to facilitate payment}. Providing a QR code encoding the payment recipient's Bitcoin address and the Bitcoin price makes payment easier for the user. QR-code generator services generally work by accepting a GET request with the data encoded as a query parameter, and returning the rendered QR code image. Thus, transaction-relevant information is contained in the GET request  (e.g., \url{https://blockchain.info/qr?data=bitcoin://[address]?amount=[price]\&size=180}). If the QR-code generator service stores HTTP requests in their logs they will have passively collected sufficient information to perform the blockchain analysis attack. We saw three domains providing this service: \url{chart.googleapis.com}, \url{qrserver.com}, and \url{blockchain.info}. Three payment processors in particular use chart.googleapis.com to generate QR codes: 
\url{coingate.com}, \url{litepaid.com}, \url{gourl.io}.\footnote{Google's policy is to retain these log for 2 weeks, for debugging and development purposes \cite{google-faq}. We could not find the retention policies for the other two service providers.}

We made purchases on a subset of 20 merchants websites, which allowed us to examine trackers on payment receipt pages.
Table~\ref{fig:confirmation-page-tps} in Appendix \ref{app:tables} presents the number of third parties found on each merchant's payment receipt page, and the number of third parties that also receive transaction-relevant information in the course of the payment flow. Embedded third parties are common on receipt pages: in total, there were \textbf{245 third parties on the 20 merchant receipt pages} we visited. 

We found further serious leaks of sensitive information on some of these pages. In particular, the payment processor Coinbase redirects to receipt page URL on the merchant website (such as \url{https://www.overstock.com/bitcoinprocessed/?...}), and appends to this merchant URL a long string of query parameters that include the Bitcoin payment address. If the resulting payment receipt page embeds third parties, then the merchant will (likely inadvertently) leak the payment address via the HTTP referer header. We found this to be the case on multiple merchant websites that use Coinbase. \textbf{The overstock.com receipt page alone leaked the payment address to 42 distinct third parties via this referer leakage.}

Additionally, we found that many merchant websites leak payment processor invoice page URLs to third parties. This is a different type of leak from the one in the previous paragraph. The URLs themselves do not  contain  sensitive information, but the contents of the invoice pages do, in the case of both Coinbase and Bitpay. In both cases, the invoice page is not protected by access control and the content can be viewed by anyone who has the URL. Of the sites from which we made purchases, 12 of the 20 merchants included leaks of these URLs to a total of 25 third parties.

\subsubsection{Third parties that receive non-BTC-denominated cart prices}
The cart page displays each product in a user's shopping cart, along with the non-BTC denominated (e.g., USD or EUR denominated) subtotal of the cart. This subtotal will often exclude taxes and shipping. 
The user is often directed to enter their shipping address at the following checkout page, which will then calculate the shipping  fee and add it to the cart subtotal.

From our crawl data we identify a second set of third parties that receive the non-BTC-denominated cart price. If the received cart price is missing shipping and handling, it will increase the adversary's uncertainty about the BTC-denominated price (see Section~\ref{sec:blockchain}). As seen in Table~\ref{fig:btc-tx-info-merchant-leak}, \textbf{43 out of the 130 bitcoin-accepting merchants we visited send some form of non-BTC-denominated cart price data to third parties} --- many more than share BTC price or address with third parties.

Based on the type of HTTP request that sent the transaction-relevant information to a third-party, we can categorize whether the sharing of data was intentional or unintentional. We consider an unintentional data leakage a sharing of data that happens solely via referer leakage.  While it is possible that the third party parses the referer for the price information on the backend, we find it is useful to separate these cases from an \emph{intentional} sharing of data. An intentional share with a third party means that the price was sent in the URL of a GET request or the body of a POST request --- in that case, the request was intentionally constructed. In our crawl, we found that \textbf{the overwhelming majority of requests that shared transaction-relevant data with a third party were intentional}: of the 312 requests we observed on 53 merchant sites sharing transaction-relevant information, 295 intentionally shared data.


To perform the cluster intersection attack, an adversary must have transaction-relevant information for at least two separate purchases. A third party positioned on more than one website is in a better position to have the necessary data. Table~\ref{fig:tx-relevant-third-party-prevalence} in Appendix \ref{app:tables} contains the prevalence of third parties receiving transaction-relevant information that appeared on at least two merchant sites. As one might expect, Google Analytics and Facebook are pervasive.



\subsubsection{Third parties receiving product page visits}
At minimum, on an e-commerce site a product page will display the non-BTC denominated price of an item that a user can purchase.
A product page will also often embed resources from many third-party domains. As an illustrative example, the product page \url{https://missionbelt.com/collections/solid-color-40mm-belts/products/vader-40}, includes resources from \emph{31} distinct third-party domains.

Thus, third-party trackers could infer the user's cart subtotal based on the product pages they visit. In our data we found several examples of third parties that not only see the product pages a user visits, but know exactly when the user adds an item to their cart. In total, \textbf{28 bitcoin-accepting merchants in our crawl shared add-to-cart events with third parties.} The most common third party to receive this information was Facebook, which received add-to-cart events on 26 merchants' sites. Being able to receive add-to-cart events is essentially equivalent to being able to see pre-shipping cart prices. Only two of those merchant sites send add-to-cart events to third parties if the consumer uses browser tracking protection.


\subsubsection{Third parties receiving transaction timing}
As discussed in Section~\ref{sec:blockchain}, an adversary needs to know the approximate timing of a Bitcoin transaction in addition to its value. There are several ways in which third parties already have this information stored in logs.

As discussed earlier, when a user completes a Bitcoin transaction, the payment processor typically redirects the user back to a receipt page. Any third parties loaded on the receipt page who had previously seen transaction-relevant data can then use the receipt page load-time as the approximate timestamp of the bitcoin transaction (except during periods of anomalously high network load). In our data, 245 distinct third-party domains had resources loaded on merchant receipt pages on sites for which we made a purchase. \textbf{16 out of 20 of those merchant sites embedded on the receipt page at least one third party which had previously received transaction-relevant information}. This knowledge makes blockchain linking much easier. A more detailed breakdown of third parties by merchant receipt page can be found in Table~\ref{fig:confirmation-page-tps} in Appendix \ref{app:tables}.

Even without seeing a bitcoin transaction receipt page, third parties could still estimate the time a transaction took place based on whether or not they know when a user starts the checkout process. From our data we measured the extent to which Facebook trackers, for instance, explicitly track a user through their API's ``\texttt{InitiateCheckout}'' event. The Facebook \texttt{InitiateCheckout} event was found on 15 sites of the 130 bitcoin-accepting merchants we visited in our crawl.

\subsubsection{Third parties receiving PII}
A leak of transaction-relevant information coinciding with a leak of PII allows the adversary to attach a real world identity to a Bitcoin address.
In our crawl, \textbf{49 bitcoin-accepting merchants leak some form of PII to a total of 137 third parties}. Table~\ref{fig:merchant-pii-counts} lists the number of merchants that share each type of PII with a third party. Table~\ref{fig:third-party-pii-counts} in Appendix \ref{app:tables} lists the number of third parties receiving each type of PII.

\begin{table}[]
\centering
\begin{tabular}{$L{2cm}^L{2cm}^L{2cm}}
\hline
\rowstyle{\bfseries} PII type & w/o protection & w/ protection \\ \hlineB{6}
email & 32  & 25 \\ \hline
firstname & 27   &  20  \\ \hline
lastname & 25  &  19 \\ \hline
username & 15 & 12 \\ \hline
address  & 13   & 9 \\ \hline
name & 11 & 4  \\ \hline
phone  &  10   & 4   \\ \hline
company & 5 & 4  \\ \hlineB{4}
\rowstyle{\bfseries} Merchants sharing PII & 49 & 38 \\ \hline
\end{tabular}
\caption{Number of merchants sharing each type of PII, with or without tracking protection}
\label{fig:merchant-pii-counts}
\end{table}

We also found that \textbf{21 third parties that receive PII also receive transaction-relevant information}. Those third parties can conduct the blockchain analysis attack and add a label to the resulting cluster.


\subsection{Other adversaries}
\label{sec:other-adversaries}

Some entities do not yet collect the necessary information to conduct the blockchain attack, but could be in the position to collect the information through an active attack.

\subsubsection{Active attacks by third parties}
Third-party JavaScript has access to the complete DOM of whatever frame the script is embedded in: if that frame contains a piece of transaction-relevant data, then a script that turned malicious could go out of its way to collect the data if it wasn't already collecting it.

OpenWPM allows us to match third-party scripts to the pieces of information they can read. In short, \textbf{107 sites in our crawl grant third-party scripts access to transaction-relevant information} (Table \ref{fig:active-tx-info-readers}). The most-prevalent third party by far is google-analytics.com, which appears to be widely trusted: for example, a google-analytics.com script is often the \emph{only} script found on Coinbase's payment processing page.

\begin{table}[]
\centering
\begin{tabular}{|l|l|}
\hline
Bitcoin address & 31 \\ \hline
Non-BTC-denominated price & 104 \\ \hline
Bitcoin price        & 30   \\ \hline
Total sites leaking transaction-relevant info & 107 \\ \hline
\end{tabular}
\caption{Active attacks: Number of merchants that allow third-party script access to transaction-relevant information}
\label{fig:active-tx-info-readers}
\end{table}

Potential third-party access to PII was even more prevalent: on the 130 bitcoin-accepting merchants we crawled, \textbf{125 merchants granted third-party scripts access to some form of PII.} This included the scripts of payment processors that may not otherwise receive PII.

\subsubsection{Network adversaries}
While most of the merchants we visited in our crawl use HTTPS, some failed to do so. In our crawl, \textbf{we found 36 merchants that did not use HTTPS for crucial parts of the payment flow, including cart pages}. A network adversary could thus see the page load in cleartext, and parse transaction-relevant information or PII from the page.

\subsubsection{Payment processors}
While users are arguably aware that payment processors will receive transaction-related information in the course of a payment, they probably do not expect that payment processors receive PII, since creating an account with the payment processor is generally not required. Yet, we found that \textbf{at least 24 merchants share some form of PII with BitPay}, even though BitPay does not require that the merchant send them PII.

\subsubsection{Merchants}
Conversely, users may expect that while merchants necessarily receive PII, they may not be able to easily identify the transaction on the blockchain. But we made merchant accounts with BitPay and Coinbase, and found that they both share the full details of the Bitcoin transaction with the merchant. 

%


\section{Blockchain analysis: Estimation of linkability}
\label{sec:blockchain}
Having shown that trackers obtain payment-related information from online purchases, we now present empirical analyses of the Bitcoin blockchain to show that trackers can use this information to uniquely identify the transaction on the blockchain. We reiterate that many trackers will have no need for the techniques in this section since merchants share unique transaction-specific information with them.

\subsection{Blockchain analysis infrastructure and algorithms}

First we describe our blockchain analysis tool, BlockSci, and the algorithms that we used for the results reported throughout the rest of this paper. We plan to release BlockSci as open-source software. 

{\bf Overview.} BlockSci is an in-memory blockchain graph database and query interface. A pre-processing step converts Bitcoin's blockchain and several other Bitcoin-like blockchains into a uniform and optimized graph representation. BlockSci exposes a Jupyter notebook interface to the analyst, and analyst queries are written in Python. The interface is object-oriented for expressiveness, but the \texttt{Transaction} objects  are not instantiated in memory when the blockchain is loaded. Instead, they are instantiated ``just in time'' when accessed.

{\bf Features.} We describe some features of BlockSci that are relevant to the analyses reported in this paper. 
\begin{itemize}
\item General graph queries: for example, we can conveniently express queries of the form ``find all transactions at a distance $\leq d$ from transaction $T$ in the graph.'' 
\item  Identifying CoinJoin transactions: The CoinJoin mixing technique aims to provide unlinkability, but not unobservability. Identifying which transactions represent CoinJoins is a key part of our attacks. Our algorithm is adapted from \cite{moser2016join}; see Algorithm \ref{alg:JoinMarketIdentification}.

\item  Address clustering: There are well-known techniques for identifying address clusters belonging to a single wallet as long as mixing techniques are not used. We adapt our clustering algorithm from \cite{meiklejohn2013fistful}, but linking addresses connected by a CoinJoin transaction. This allows us to avoid minimize positives. See Algorithm \ref{alg:clustering}.

\item P2P network data collection: Some transactions broadcast on the P2P network may never be included in the blockchain (for example, if the transaction fee provided is too low). BlockSci includes a network daemon that monitors and logs all broadcast transactions. This information is necessary for the adversary to minimize transaction time uncertainty (Section \ref{sec:threatmodel}). Note that the adversary doesn't necessarily need to collect the data himself; broadcast timestamps for all historical transactions are available from public sources such as blockchain.info.

\item  Exchange rate data collection: BlockSci incorporates exchange rate data for various cryptocurrency / fiat currency pairs of interest, and exposes them via the same interface. This information is also necessary for the adversary in our attacks.

\item Support for altcoins: BlockSci supports a range of blockchains, which are currently  Bitcoin, Namecoin, Litecoin, and Dash. It is straightforward to support at least the eight other Bitcoin-derived blockchains supported by the custom parsing tool that we use. Blockchains with complex scripts, notably Ethereum, are out of scope. Most of our analyses in this paper were done on the Bitcoin blockchain, but we do present an analysis for Litecoin (Section \ref{sec:blockchain-results}).

\end{itemize}

\subsection{Method} 

We seek to answer the question: for a given level of price uncertainty, exchange rate uncertainty, and transaction time uncertainty, what is the distribution of anonymity set sizes of the transaction? The anonymity set contains candidate transactions, one of which represents the actual payment. Based on the anonymity set sizes, we also compute adversary success probabilities. Throughout, we average our measurements over a set of prices (obtained from actual sites, as described below), and a set of random points in time over a two-year period from mid-2015 to mid-2017.

{\bf Exchange rate data.} Recall from Section \ref{sec:background} that payment processors use price data from exchanges, which is also available to adversaries. In our measurements, we use publicly available historical data from BitStamp made available by bitcoincharts.com.\footnote{\url{https://api.bitcoincharts.com/v1/csv/}} The data contains the prices of all trades executed on the exchange, starting in September 2011. As of June 2017 it contains 11.6 million trades. During the time period of interest to us it contains about 4.4 million trades, or about 4.2 trades per minute.

{\bf Sampling prices.} To obtain a representative sample of prices of user purchases, we sampled 100 item prices from our dataset of merchants. We sampled merchants randomly and then sampled items randomly from those listed on the homepages of those merchant websites, ensuring a maximum of 10 from any one merchant. When converted to USD, the prices ranged from a minimum of 1.52 to a maximum of 359.00, with a mean of 51.27 and a median of 24.99. 

Sampling actual prices is important, because the distribution of values of e-commerce payments is different from that of other transactions on the blockchain. For example, prices are often close to integer multiples of the currency of account (USD, EUR, etc.) \cite{gervais2015quantifying}. Therefore, if we sample prices directly from blockchain transactions, we might obtain unrepresentative results.


{\bf Sampling times.} We pick 100 random timestamps from our time period of interest. By picking these randomly instead of periodically, we ensure that there are no patterns such as specific times of day or day of week.

\begin{figure}
\centering
\includegraphics[width=0.7\columnwidth]{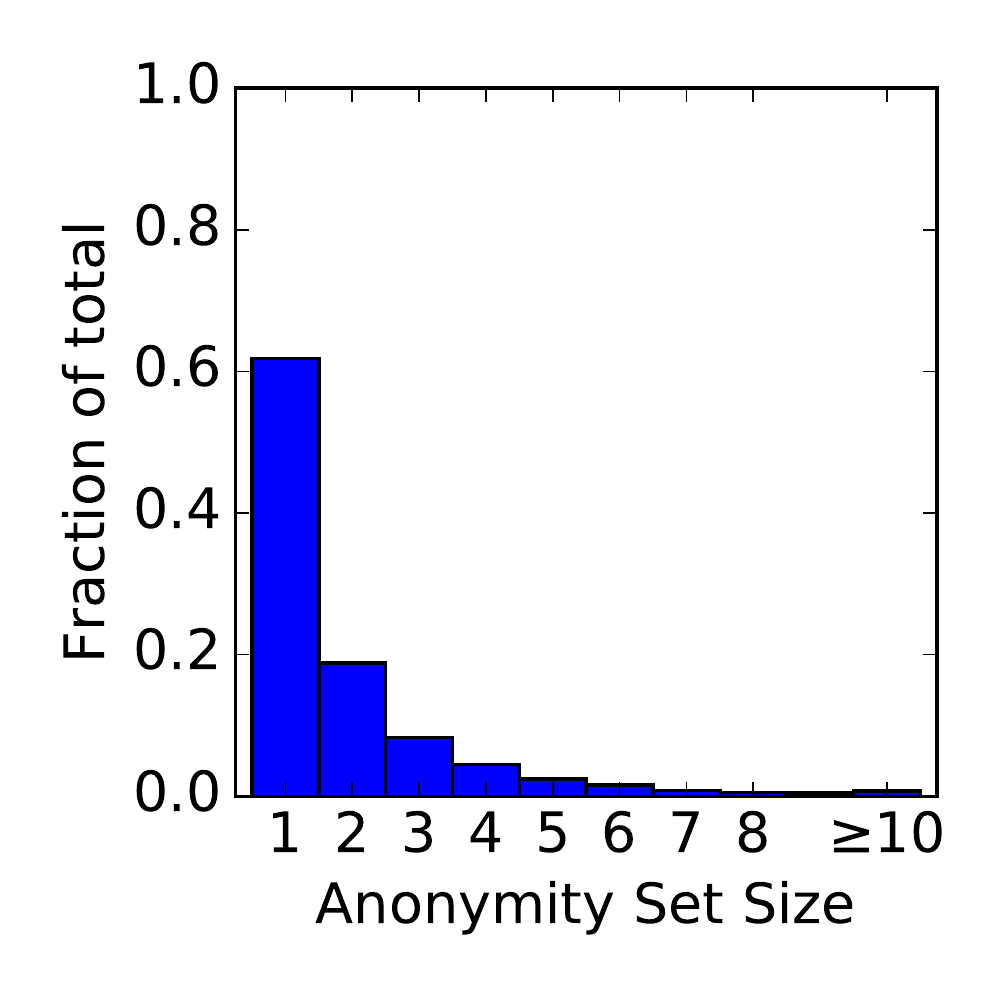}
\caption{Anonymity set size of the single-transaction linkage attack that aims to link a web transaction to the blockchain. Here the price uncertainty set size is 5, the payment time uncertainty is 15 minutes, and the exchange rate uncertainty is 5 minutes.}
\label{fig:anonymity_set}
\end{figure}

\begin{figure*}
\centering
\begin{minipage}{.33\textwidth}
\centering
\includegraphics[width=.9\columnwidth]{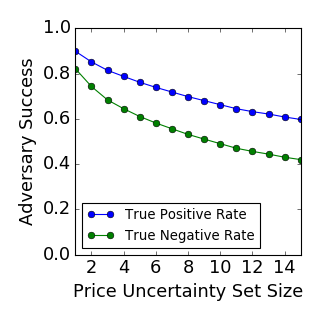}
\caption{Effect of price uncertainty on transaction linkage}
\label{fig:price_uncertainty}
\end{minipage}\hfill
\begin{minipage}{.33\textwidth}
\centering

\includegraphics[width=.9\columnwidth]{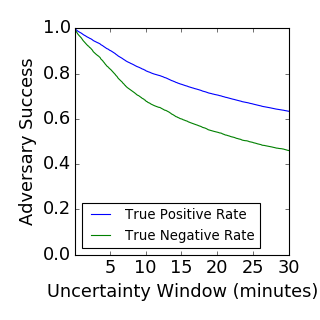}
\caption{Effect of payment time uncertainty on transaction linkage}
\label{fig:payment_time_uncertainty}
\end{minipage}\hfill
\begin{minipage}{.33\textwidth}
\centering
\includegraphics[width=.9\columnwidth]{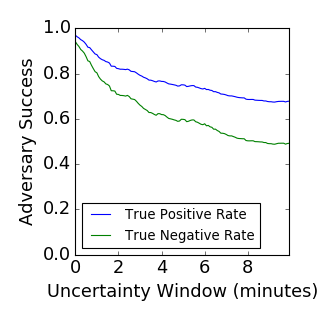}
\caption{Effect of exchange rate uncertainty on transaction linkage}
\label{fig:exchange_rate_uncertainty}
\end{minipage}
\end{figure*}

{\bf Modeling price uncertainty.} As we showed in Section \ref{sec:webmeasurement}, the adversary typically sees the USD (or other local currency) price in the shopping cart, but only some trackers see the final price after shipping and any taxes. By analyzing the behavior of various merchants, we make the following key observation: {\em if the adversary knows only the country of the user's shipping address, there are only a small number of possible values (typically fewer than 10) for the difference between the cart price and the final price.} For example, there are no instances where each US state has a distinct shipping fee. Based on our observations, the vast majority of merchants do not collect tax. Note that often the shipping address is directly revealed to the tracker (Section \ref{sec:webmeasurement}). Even otherwise, it is easy for trackers to learn or infer the user’s country, whether based on IP address, locale, or language. Thus, we model price uncertainty as a small set of possible values. We sample these values from the list of actual shipping rates on the merchant websites we analyzed. We always include the value 0 in this sample, because free shipping is (unsurprisingly) a popular option.

{\bf Modeling time uncertainty.} The tracker may observe the user take one or more of the following actions: view the shopping cart, initiate checkout, and view the transaction receipt page. In the first case, the tracker can guess that the user may have initiated payment within a few minutes (of course, this guess might be incorrect). In the second case, the tracker knows that if the payment was made at all, it would have to be within a time window set by the payment processor, typically 15 minutes. In the third case, the tracker knows the transaction timestamp to within a few seconds (network latency).

{\bf Modeling exchange rate uncertainty.} The exchange rate data used by the payment processor is always available to the adversary. However, there is potentially some uncertainty introduced by the lag between the tracker observing the user initiate the checkout process and the user being actually quoted an exchange rate. We model the adversary's uncertainty as a time interval. If this window is 5 minutes, it means that the exchange rate that was applied to the transaction could take any value from the published time-series of exchange-rate values during a 5-minute period that begins when the adversary observed the user initiate checkout. 

{\bf Modeling the victim and adversary.} We simulate 10,000 payment flows based on all combinations of the 100 prices and 100 timestamps sampled as above. For each flow, we consider two cases: the victim either does, or doesn't, complete the payment within the windows of uncertainty assumed by the adversary.

We posit an adversary that behaves as follows: 
\begin{itemize}
\item
if there is exactly one transaction that falls within the uncertainty intervals, output that transaction.
\item
if there is more than one such transaction, output a random transaction from that set.
\item
if there are no such transactions, output ``no such transaction''.
\end{itemize}

The last point is important, because in many cases the adversary observes the user on the shopping cart page or the checkout page, but does not know for sure that the payment was made (some adversaries are also present on transaction receipt pages and will have this additional information). 

{\bf Quantifying success.} We quantify the adversary's success in terms of two numbers: the true positive rate and the true negative rate. A true positive occurs when the victim completed the transaction, and the adversary outputs the correct transaction. A true negative occurs when the victim did not complete the transaction, and the adversary correctly outputs ``no such transaction''.

For each simulated payment flow and each set of uncertainty parameters, we search the log of broadcast transactions (``mempool" log) for transactions that match the price and time within the specified uncertainty windows. Any transactions found, {\em in addition to the payment itself}, constitute the anonymity set. Naturally, since the payments are simulated, we do not expect to find them on the blockchain, but in a real attack the payment would be part of the blockchain. In other words, our measurements answer the question: ``had a payment of a given value been made at a given time, how many transactions on the blockchain could it be confused with?'' The anonymity set size is 1 more than this value.

{\bf Additional heuristics.} E-commerce payment transactions have several other characteristics that enable the adversary to distinguish them from (some) other transactions on the blockchain. We incorporate several such heuristics in our attack.

\begin{itemize}
\item Payments are always made to regular addresses rather than high-security ``multisignature'' addresses. This is true across almost all 130 merchants that we analyzed. The use of multisignature addresses would make our attack far stronger since the attacker, knowing the type of address used by any given merchant, would be able to greatly limit the set of candidate transactions on the blockchain.

\item Payment transactions almost always have two outputs --- the recipient's output and the change output --- and never more than two. This behavior is consistent across all but one user wallet software that we are aware of; the exception is Samourai Wallet (\url{https://samouraiwallet.com/}).

\item Fresh addresses are used, both for change and for the recipient's output. This is a conservative assumption; alternative behavior would make our attack stronger. If the user's wallet reuses addresses for change, that would undo the effect of mixing. If the recipient reuses addresses, it would make it easier for the adversary to associate specific addresses with recipients, and thus further filter the set of candidate transactions on the blockchain.

\end{itemize}

\subsection{Results}
\label{sec:blockchain-results}

{\bf Anonymity set size.} Figure \ref{fig:anonymity_set} shows the distribution of anonymity set sizes under default values of various parameters: payment time uncertainty of 15 minutes, exchange rate uncertainty of 5 minutes, and a price uncertainty set size of 5. The  most common value of the anonymity set is 1, which shows that the attack is powerful under this default set of parameters. Based on the anonymity set size distribution, the true positive rate is 76\% and the true negative rate is 62\%.

{\bf Impact of uncertainty.} Having shown that the attack is successful under a default set of parameters for uncertainty, we examine the impact of each uncertainty parameter. In Figure \ref{fig:price_uncertainty} we see that the accuracy remains high even if there are 10 possible values for the price. As we observed earlier, price uncertainty arises due to shipping options, and there are rarely more than 10 possible values for it for any given country. 

Similarly, the attack degrades gracefully when we increase the adversary's time uncertainty or exchange rate uncertainty (Figure \ref{fig:payment_time_uncertainty}, Figure \ref{fig:exchange_rate_uncertainty}). Note that if the payment processor automatically redirects to the payment receipt page, and the adversary is embedded on this page, then the time uncertainty is on the order of seconds, and the success rate is extremely high. 

More generally, the adversary will have a high success rate if his uncertainty on {\em at least one} of the three dimensions is low (Figures \ref{fig:price_uncertainty}, \ref{fig:payment_time_uncertainty}, \ref{fig:exchange_rate_uncertainty}), as this greatly cuts down the number of possible matching transactions.

{\bf Robustness of the results.} While we took care to sample prices from the actual distribution of prices on merchant websites, we find that our results are robust in terms of the sampling strategy. For example, we repeated our experiments with prices sampled from the distribution of transaction amounts on the blockchain (Figure \ref{fig:payment_time_uncertainty_bitcoin} in Appendix \ref{app:figures}). The results are very similar; the accuracy improves slightly. We also repeated our experiments with all prices doubled, i.e., with each sampled price replaced by twice its value (Figure \ref{fig:payment_time_uncertainty_doubled} in Appendix \ref{app:figures}). Again the results are essentially unchanged.

We also repeated our experiments on the Litecoin blockchain instead of Bitcoin. Litecoin is the original altcoin --- the first fork of Bitcoin --- and is the altcoin with the most adoption for online payments, in terms of support by merchants and payment processors. Again we find that the success rate is high (Figure \ref{fig:payment_time_uncertainty_litecoin} in Appendix \ref{app:figures}); in fact, it is higher than the success rate for Bitcoin, likely due to Litecoin's lower transaction volume, and therefore smaller anonymity sets. Litecoin had a volume of 3,605,028 transactions in the two-year period of interest, as opposed to Bitcoin's 150,614,721.

{\bf Further improvements.} So far, we have made conservative assumptions about the adversary's knowledge. The success of the attack in practice may in fact be much higher, either due to idiosyncratic behavior by payment processors or due to additional information available to the adversary.

BitPay, one of the two main payment processors, rounds its transaction amounts (in Satoshis) to a multiple of 100. Since the adversary knows the identity of the payment processor, whenever that processor is BitPay, he can eliminate a large fraction of possible transactions --- any transaction amount that is not a multiple of 100 Satoshis can be eliminated. Applying this heuristic, the accuracy improves substantially (Figure \ref{fig:payment_time_uncertainty_bitpay} in Appendix \ref{app:figures}).

Even if there is no discernible pattern in the transaction amount, the adversary may be able to tell which (if any) payment processor was involved in any given transaction on the blockchain. Such address tagging heuristics are well known \cite{meiklejohn2013fistful}, and are applied at scale by companies such as Chainalysis. Tagging is not always accurate, but it can help the adversary greatly decrease the anonymity set. This technique was used for Bitcoin forensics in a recent paper \cite{portnoff17backpage}.

\section{The cluster intersection attack}
\label{sec:cluster}
We now turn to our second attack, the cluster intersection attack (Algorithm \ref{alg:cluster-intersection}). To recap, the attack is applicable when the adversary has auxiliary information revealing that two (or more) transactions made with mixed coins trace back to the same wallet (address cluster). Web trackers who observe multiple purchases may have this information.

\begin{algorithm}
\begin{algorithmic}[1]

\Function{ExpandCluster}{$addr$}
  \State $C \gets \{addr\}$
  \ForAll{$tx$ in $\Call{TxsFrom}{a}$}
    \If{not $\Call{IsMixTx}{tx}$}
      \State $C \gets C \cup \Call{FromAddresses}{tx}$
	  \State $C \gets C \cup \{\Call{ChangeAddress}{tx}\}$
    \EndIf
  \EndFor
  \ForAll{$tx$ in $\Call{TxsTo}{a}$}
    \If{not $\Call{IsMixTx}{tx}$ and \\
    \ \ \ \ \ \ \ \ \ \ \ \ \ \ \  $\Call{ChangeAddress}{tx} = addr$}
      \State $C \gets C \cup \Call{FromAddresses}{tx}$
    \EndIf
  \EndFor
  \Return $C$
\EndFunction
\caption{{\bf Clustering.} One step of the address clustering algorithm. We invoke this function recursively to find all addresses associated with a given coin or address. The algorithm incorporates the multi-input and change-address detection heuristics from \cite{meiklejohn2013fistful}. Bitcoin mixing today is dominated by JoinMarket, so we use JoinMarket detection (Appendix \ref{sec:joinmarket-identification}) in place of IsMixTx.} \label{alg:clustering}
\end{algorithmic}
\end{algorithm}

\begin{algorithm}
\caption{{\bf Cluster Intersection Attack.} Step 1 can be amortized over multiple invocations of the algorithm; alternately step 2 can be modified so that join detection can be performed only as needed.}\label{alg:cluster-intersection}
{\em Inputs:} 
\begin{itemize}
\item a set of mixed coins $C$ known to be controlled by the same user.
\item an integer $r$, representing the adversary's (possibly incorrect) assumption that the victim did at most $r$ rounds of mixing.
\end{itemize}

{\em Output:} a wallet cluster.

\begin{enumerate}
\item Identify all join transactions on the blockchain.
\item For each coin $c \in C$:
\begin{itemize}
\item Identify all coins $x$ such that there is a directed path from $x$ to $c$ of length at most $r$ consisting only of join transactions. Call this set of coins $X_c$.
\item For each coin $x \in X_c$, identify the wallet cluster it belongs to (Algorithm \ref{alg:clustering}). Call the resulting set of wallet clusters $W_c$.
\end{itemize}
\item Compute the intersection of wallet clusters: $\bigcap_{c \in C} W_c$. 
\item If it results in a unique wallet cluster, output it. Otherwise output ``incorrect assumptions''.
\end{enumerate}
\end{algorithm}

In this section, we present a large-scale simulation of the effectiveness of this attack. An empirical validation, where we de-anonymize our own wallets, is deferred to the next section.

\subsection{Method}

\begin{figure*}
\centering
\begin{minipage}[t]{.49\textwidth}
\centering
\includegraphics[width=.714\columnwidth]{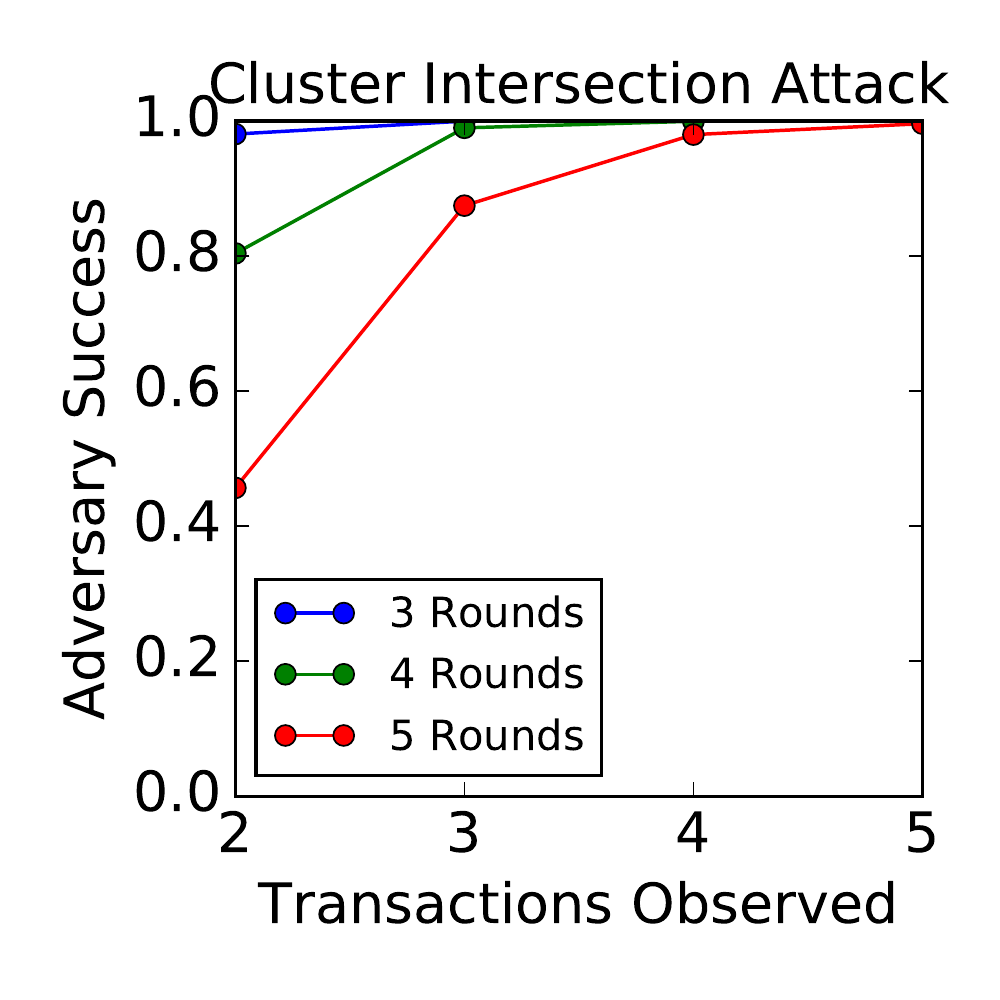}
\caption{Effect of the number of rounds of mixing on  success rate of cluster intersection attack. For $r < 3$ rounds, the success rate is 100\%.}
\label{fig:jm_observations}
\end{minipage}\hfill
\begin{minipage}[t]{.49\textwidth}
\centering
\includegraphics[width=.714\columnwidth]{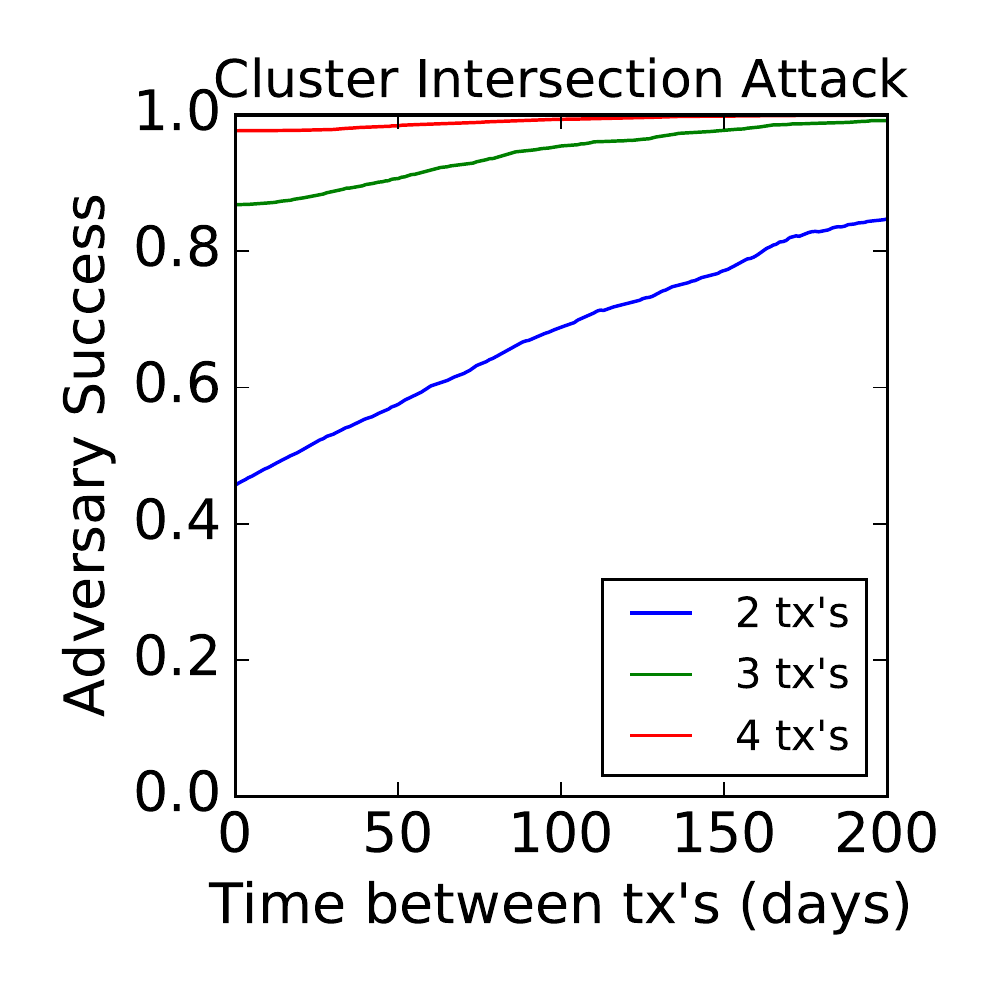}
\caption{Effect of age of mixed coins on success rate of cluster intersection attack (with 5 rounds of mixing). The X axis is the maximum difference in age between coins known to the adversary to be associated with the same user.}
\label{fig:jm_time}
\end{minipage}
\end{figure*}

{\bf Identifying joins.}
First we identify existing CoinJoin transactions on the Bitcoin blockchain. We focus on JoinMarket, since it is (to our knowledge) the only decentralized mixing service that is currently operating and has a usable level of liquidity. We adapt M\"oser et al.'s algorithm to identify JoinMarket transactions \cite{moser2016join}, and it is shown in Algorithm \ref{alg:JoinMarketIdentification}. We found 95,239 such transactions, of which 78,697 are during the period of interest to us (mid 2015--mid 2017). The number of coins mixed in one of these transactions has a mean of 3.98 and a standard deviation of 1.72.

{\bf Simulating the victim.} 
We consider a victim with a wallet of clusterable addresses who obtains 100 distinct mixed coins over the two-year period of interest. We sample 100 timestamps (block heights) uniformly during this period; at each of these times, the victim initiates mixing of a coin from her wallet and completes $r \leq 5$ rounds of mixing. 5 rounds represents a very high degree of anonymity based on JoinMarket's advice to users \cite{joinmarkettumblerguide}. The victim retains the mixed coins until the end of the period of interest. The values of these coins don't matter since this information is not used by our deanonymization algorithm.

To simulate the mixing of one coin (with $r$ rounds of mixing) starting from a given timestamp (block height), we sample from the JoinMarket transactions on the blockchain that have this timestamp. With this node as the starting point in the graph of JoinMarket transactions, we sample a path of length $r$ from among all such paths. If there are no such paths, we repeat the procedure starting from a different initial transaction.

{\bf Attack.} At this point the victim has 100 mixed distinct coins in her (simulated) wallet. Now we simulate the web tracker's view, that is, we simulate the victim  making two transactions in a way that reveal to the adversary that two of these coins trace from the same wallet. Then we execute the cluster intersection attack (Algorithm \ref{alg:cluster-intersection}). We repeat the procedure with different values of the number of rounds $r$ and the number of transactions $t$ observed by the adversary.

\subsection{Results}
Figure \ref{fig:jm_observations} shows the adversary's success rate as a function of the number of rounds of mixing and the number of transactions observed by the adversary. By construction of the experiment, the cluster intersection attack has the same true positive rate and true negative rate. Thus the graph also represents the probability that, if the adversary's assumptions are incorrect about the number of rounds of mixing, it will output ``incorrect assumptions''. With one or two rounds of mixing, just two observed transactions are sufficient for the adversary to identify the wallet cluster. Even with four rounds of mixing, a small number of observations is sufficient for high accuracy. 

Figure \ref{fig:jm_time} helps explain why the attack succeeds: the success rate is strongly dependent on the difference in age between the different mixed coins. This is intuitive: if the victim mixed a coin a year ago and another coin today, the anonymity sets of the two coins are much less likely to intersect, compared to two coins both mixed today. In other words, users who have a long history of making e-commerce purchases using mixed coins are at a greater risk of deanonymization, not just because of the number of purchases but also because of the gap between them.

\section{Empirical validation of attacks}
\label{sec:end-to-end}
In this section we describe how we validated our attacks empirically by making actual purchases and participating in CoinJoin transactions. Naturally, the scale of these experiments was more limited than our previous measurements.

\subsection{Setup.} We began by purchasing bitcoins from exchange service Coinbase and routing it to a set of six addresses. We ensured that these six addresses are clearly clusterable by our clustering algorithm (Algorithm \ref{alg:clustering}). This simulates a user with a wallet containing addresses that are linkable to each other, before employing mixing. As discussed in Section 2, we believe that this is a conservative and realistic assumption. Furthermore, the clusterability of the user's wallet is affected by factors not in the user's control. For example, if a payment processor provides a payment address that has been used before, rather than a freshly generated address, then the user's change address will be linked to her wallet. 

Next, we participated in JoinMarket CoinJoin transactions to create 11 coins which are not linkable to the main cluster using known techniques. We participated in one round of CoinJoin for 6 of the transactions and two rounds of CoinJoin for the other 5.


Finally, we made a set of 21 purchases on 20 merchant sites. 
We sampled these sites from among those that leaked transaction-relevant information to at least one tracker (as measured in Section \ref{sec:webmeasurement}). For 11 of these purchases, we used coins that had been mixed in the previous step, and we ensured that these addresses as well as the change addresses for these purchases did not get linked to our cluster. For the other 10 purchases, we used coins directly from our cluster. The final prices of these items ranged from a minimum of 3.28 to a maximum of 46.40 when converted to USD, with a mean of 13.67.




\subsection{Validating transaction linkability.}

\begin{figure}[t]
\centering
\includegraphics[width=0.7\columnwidth]{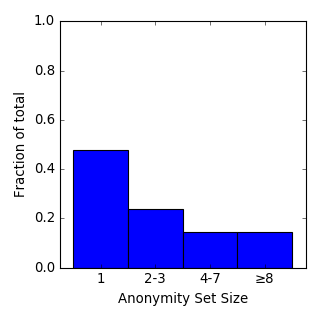}
\caption{Observed anonymity set sizes in empirical evaluation of the transaction linkage attack}
\label{fig:e2e_anonymity_set}
\end{figure}

We calculate the anonymity sets of these 21 transactions based on our default values of the adversary's uncertainty: a 5-minute exchange rate uncertainty and a 15-minute payment time uncertainty. The uncertainty windows are centered around the true values of the payment time and the exchange rate determination time. As for price uncertainty, we use the actual list of shipping options (and resulting list of final prices) that we recorded while making purchases. The number of possible pre-BTC prices is typically 5 or fewer per purchase. We find that in 10 out of 21 cases (48\%), the anonymity set size is 1 (Figure \ref{fig:e2e_anonymity_set}).

For 17 of the 21 purchases, there was at least one tracker that received transaction-relevant information and was also present on the payment receipt page, which means that for these adversaries the payment time uncertainty is very low. In this scenario, the anonymity set sizes are much lower: in fact, it is 1 for 14 of 17 purchases. The adversary behaving as described in Section \ref{sec:cluster} would identify the correct transaction 90\% of the time. The true negative rate is also high: 82\%. This means that if the adversary's assumption about the payment time was incorrect for whatever reason --- if the P2P network load is too high or the user's wallet software included too low a transaction fee, the receipt may not happen quickly --- the adversary will be able to correctly deduce this.

\subsection{Validating the cluster intersection attack.}

Next we validate the cluster intersection attack. Out of the 11 purchases we made using mixed coins, we consider adversaries that observe a random $t$ of those purchases and know which transaction on the blockchain corresponds to each of them. We find that for $t=2$, the adversary described in Algorithm \ref{alg:cluster-intersection} has an 89\% chance of correctly identifying our wallet cluster, and for $t=3$, this goes up to 99\%.

If more rounds of mixing are used, the attack will not work as well. On the other hand, we mixed all our coins during a 3-day period, and realistic users who mix coins over a period of months or years will have worse privacy. In Section \ref{sec:blockchain} we evaluated the effect of these factors as well as the number of transactions observed by the adversary. For our experiments in this section, we limited the number of parameters because of the expensive nature of the mixing market. 

\subsection{Validating the attack end-to-end.}
The two attacks are especially powerful when combined. Even if the single-transaction linkage doesn't produce a unique transaction, we can run cluster intersection starting from every possible combination of the candidate transactions produced by it. Most combinations will produce an intersection of size zero, and can be discarded. If exactly one combination produces one cluster and all the rest produce zero clusters, then the adversary outputs that cluster.

From our purchase records, we determined that there were 11 trackers in a position to observe more than one purchase: americanexpress.com, chatid.com, criteo.com, doubleclick.net, facebook.com, google-analytics.com, google.com, monetate.net, revjet.com, steelhousemedia.com, tealiumiq.com. Overall there are 25 pairs of purchases for which there was a tracker that could observe both purchases. We ran the end-to-end attack on all 25 pairs, again using default values of the adversary's payment time uncertainty and exchange rate uncertainty. We found that the attack succeeds in identifying our wallet cluster in 20 cases.

Again, these numbers reflect conservative assumptions about the adversary's knowledge. If the Bitcoin transaction details are directly leaked to the tracker, or if the tracker is present on the receipt page, or observes more than two payments, the success rate will be much higher. Similarly, many Bitcoin users do not employ mixing. We do not know what fraction of e-commerce purchases are made with mixed coins, but we observe that only .05\% of transactions on the blockchain over the past two years are CoinJoin transactions. If the user doesn't employ mixing, then it will be straightforward for the adversary to identify her wallet cluster, even with a high degree of uncertainty in the payment amount and time.

\section{Mitigation and discussion}
\label{sec:mitigation}

Our findings are a reminder that systems without provable (or at least well-defined) privacy properties may have information leaks and privacy breaches in unexpected ways. When multiple such systems interact, the leaks can be even more subtle. For another example of the difficulty of composing systems with complex privacy properties, see \cite{biryukov2015bitcoin}. 

Cryptocurrency anonymity is a new research topic, but it sits at the intersection of anonymous communication and data anonymization, both well-established fields. Unfortunately, it seems inherit the {\em worst} of these two worlds. Like data anonymization (and unlike anonymous communication), sensitive data must be publicly and permanently stored, available to any adversary, and de-anonymization may happen retroactively. And like anonymous communication systems (and unlike data anonymization), privacy depends on subtle interactions arising from the behavior of users and applications. Worse, realistic traces of the system may not be available at the time of designing and implementing the privacy defenses.

Turning to defenses, we observe that our first attack exploits the inherent tension between privacy and e-commerce, and our second attack exploits the inherent tension between privacy and the public nature of the blockchain. Thus, all mitigation strategies come with tradeoffs. The available mitigations break down into three categories: self-defense by users, techniques that merchants can use, and alternative cryptocurrencies or cryptocurrency-based payment methods. 

{\bf Mitigation by merchants.} There are a few straightforward mitigations that merchants could deploy: (1) enabling HTTPS on all shopping (and especially payment-related) pages --- this would protect against network adversaries, but not third-party trackers, our main adversary of interest (2) generating Bitcoin-address QR codes internally instead of outsourcing it to a third party; (3) avoiding leaks of the Bitcoin address from payment receipt pages; and (4) avoiding unintentional PII leaks. As to the last point, however, note that the attack succeeds as long as {\em some} first party website visited by the user leaks PII to third parties, and at least some PII leakage is for cross-device linking purposes \cite{xdt}, and thus intentional. Beyond these obvious steps, merchants could share less data with third parties, and with fewer of them, but this would come at the expense of their advertising and analytics objectives.

{\bf Self-defense.} Web tracking is a well-known privacy threat, and the main defense is to use browser extensions such as uBlock Origin, Adblock Plus, or Ghostery to block trackers. Such defenses can be quite effective, but they are far from perfect \cite{englehardt2016online,merzdovnik2017block}. Our measurements in Section \ref{sec:webmeasurement} confirm the partial effectiveness of these tools. Note that these tools are of no help when the adversary is a network eavesdropper (for either attack) or the payment processor (for the cluster intersection attack).

On the cryptocurrency side, the main self-defense is to use improved mixing techniques, especially multi-round mixing. We showed in Section \ref{sec:cluster} that this is effective (but not perfect) as long as the adversary observes only 2 or 3 transactions. In our end-to-end evaluation in Section \ref{sec:end-to-end}, we carried out only 1 or 2 rounds of mixing, and this a limitation of our experiments. Increasing the number of rounds comes at the expense of cost (transaction fees and mixing fees) and convenience (due to transaction confirmation time). A more through evaluation of the trade-offs is a topic for future work. 

While we have focused on CoinJoin or decentralized mixing, in principle the cluster intersection attack should also work against centralized mixes. If a mixing service introduces a delay of (say) up to 6 blocks, then for a given coin that was mixed at a given block height, all mix outputs produced in the next 6 blocks can be considered part of its anonymity set. The main complication is the extent to which mix transactions are distinguishable from other transactions, which is likely highly implementation-dependent. Of course, centralized mixing is additionally vulnerable to the mix operator colluding with the adversary or stealing the funds.  Evaluating the attack against centralized mixes (as well as other anonymity techniques including TumbleBit \cite{heilman2016tumblebit}) is an avenue for future work.


{\bf Alternative cryptocurrencies and payment mechanisms.} Unlike the Bitcoin approach of anonymity as an overlay, other cryptocurrencies bake privacy into the protocol, and promise untraceability of transactions. The most well known of these are Zcash, based on the Zerocash protocol \cite{miers2013zerocoin,zerocash}, and Monero, based on the Cryptonote protocol \cite{van2013cryptonote}. Zcash is more computationally expensive but comes with more rigorous security properties. Of the two, Monero has more vendor support at the time of writing, but still far less than Bitcoin or even Litecoin, and primarily on hidden-service sites merchandising illicit goods. While some anonymity weaknesses have recently been revealed in Monero \cite{miller2017empirical,kumar2017traceability}, we believe that it is not susceptible to the cluster intersection attack.

The lightning network \cite{poon2015bitcoin} is a proposal for a fast micropayments. It is a network of two-party bidirectional payment channels. If Alice wants to pay Bob, she finds a path of such channels that link her to Bob, through which she can route the payment. Although the lightning network relies on Bitcoin (or another underlying cryptocurrency) for its security, the vast majority of transactions are off-chain. There is no global ledger of all lightning payments, rendering our attack ineffective. If and when the lightning network is deployed on Bitcoin, it would be an effective defense. However, other privacy concerns have been identified \cite{atlas2017inevitability,malavolta2017concurrency}, and the issues that arise are analogous to communications anonymity \cite{herrera2016privacy}. 


Finally, like virtually all deanonymization attacks on cryptocurrencies, our techniques could be used to build forensic tools for law enforcement use. In past investigations, agents have sought to find the identity behind specific blockchain transactions that were known to represent thefts, funding of unlawful activities, or earnings from unlawful activities, as in the case of ransomware. Alternatively, agents may have an identified person of interest and may wish to scrutinize their cryptocurrency dealings for evidence of money laundering or other financial crimes. Thus, both blockchain $\rightarrow$ web and web $\rightarrow$ blockchain linking techniques are of potential interest to law enforcement. Agents might subpoena a tracker or payment processor for information that might allow such linkage, or even use network surveillance techniques.

{\bf Acknowledgements.} We are grateful to Malte M\"oser and Ben Burgess for feedback on a draft. This work is supported by NSF awards CNS-1421689, CNS-1526353, CNS-1651938, and an NSF Graduate Research Fellowship under
grant number DGE-1148900.

\bibliographystyle{IEEEtran}
\bibliography{btctracking}

\clearpage
\appendix
\section{Appendix}
\label{sec:appendix}

\subsection{JoinMarket identification}
\label{sec:joinmarket-identification}

CoinJoin transactions have a distinct structure, and JoinMarket transactions especially so. Whether or not JoinMarket can be modified to operate in a way that the transactions are not as distinguishable from other transactions is an open question. Here we describe our algorithm for identifying JoinMarket transactions, and evaluate its effectiveness. We adapt the algorithms from several previous works \cite{moseranonymous,moser2016join,selij2015coinshuffle,atlas2014coinjoin,meiklejohn2015privacy }.

\begin{algorithm}[H]
\begin{algorithmic}[1]
\Function{IsJoinMarketTransaction}{$tx$}
  \If{\Call{ContainsOpReturn}{tx}}
  	\State \Return false
  \EndIf
  \State $p \gets \ceil{|\Call{Outs}{tx}| / 2}$
  \If{$p < 2$}
  	\State \Return false
  \EndIf
  \State $v \gets \Call{MostCommon}{\Call{Val}{O} : O \in \Call{Outs}{tx}}$
  \If{$|\{O \in \Call{Outs}{tx} \mid \Call{Val}{O} = v\}| \neq p$} 
  	\State \Return false
  \EndIf
  \State $A \gets \{\Call{Addr}{x} : x \in \Call{Ins}{tx}\}$
  \State $V \gets \{\}$
  \ForAll{$a$ in $A$}
  	\State $s \gets 0$
  	\ForAll{$I$ in $\Call{Ins}{tx}$}
    	\If{$\Call{Addr}{I} = a$}
        	\State $s \gets s + \Call{Val}{I}$
        \EndIf
    \EndFor
    \State $V \gets V \cup \{s\}$
  \EndFor
  \State $q \gets \Call{MaxFee}{v}$
  \State $B \gets $ Array of length $p$ with all entries $v - q$
  \State $i \gets 0$
  \ForAll{$O$ in $\Call{Outs}{tx}$}
  	\If{$\Call{Val}{O} != v$}
    	\State $B[i] \gets B[i] + \Call{Val}{O}$
        \State $i \gets i + 1$
    \EndIf
  \EndFor
  
  \ForAll{$P$ in $\Call{Partitions}{V}$}
  	\State $t \gets 0$
  	\ForAll{$S_i$ in $P$}
    	\If{$\sum S_i \geq B_i$}
        	\State $t \gets t + 1$
        \EndIf
    \EndFor
    \If{$t \geq p$}
    	\State \Return True
    \EndIf
  \EndFor
  \State \Return False
\EndFunction
\end{algorithmic}
\caption{{\bf JoinMarket Identification.} }
\label{alg:JoinMarketIdentification}
\end{algorithm}

The algorithm is a series of heuristics to filter transactions based on the following observations:
\begin{itemize}
\item Transactions should only contain spendable addresses (lines 2--3)
\item There must be at least two participants. If there are $n$ participants, there could be either $2n$ or $2n - 1$ outputs because JoinMarket has “sweep transactions” where the taker obtains no change (lines 4--6)
\item There must be an output of value $v$ for each participant, $v$ being the mot common output value (lines 7--9)
\item There must be enough inputs to cover all of the outputs (lines 10--32). Specifically, for each change address, there must be a distinct set of inputs that add up to at least the output value $v$ plus the change value minus the max fee ($q$) that might have been paid to the liquidity providers. For our calculations we set this to be the maximum of .0001 satoshis or 1\% of the CoinJoin output.
\end{itemize}

One limitation of this algorithm is that it is slow when the number of inputs is large. This is unavoidable as the problem is NP-complete (variable-sized bin-covering in the unit supply model \cite{hellwig2012approximation}). The listing shows a brute-force implementation for simplicity; our actual implementation is optimized, but nevertheless exponential. For our analyses, we ran it on transactions with at most 17 inputs; it takes about 30 minutes to process 150 million transactions. Based on the work of M\"oser et al., who don't use this heuristic, the vast majority --- 92\% --- of JoinMarket transactions have no more than 17 inputs. 

When we run this algorithm on our two-year period of interest (May 2015 -- April 2017; block 354416--block 464269), it results in 78,697 transactions. The algorithm has low false negatives, and thus we regard this as a near-superset of JoinMarket transactions for this period. The criteria used in Algorithm \ref{alg:JoinMarketIdentification} for filtering transactions are necessarily true of all JoinMarket transactions, except for any transactions where liquidity providers charged so high a fee that they were rejected by our max-fee heuristic. But based on the empirical analysis of \cite{moser2016join}, virtually all offers posted on the market by makers have a fee that is significantly less than the threshold we used. As a further sanity check, all CoinJoins that we performed in our experiments (Section \ref{sec:end-to-end}) are identified by this algorithm.

CoinJoins and especially JoinMarket transactions tend to connect to each other, and thus we can expect to find large connected components among the identified transactions. Indeed, among the 78,697 transactions, we find a single giant component of size 60,187. We regard these as a near-subset of CoinJoin transactions during this period. While it is possible that non-CoinJoin transactions may sometimes accidentally satisfy the criteria in Algorithm \ref{alg:JoinMarketIdentification}, it is unlikely that they will cluster with the JoinMarket transactions. 

The fact that our near-superset and our near-subset are similar in size gives us further confidence in the method. Depending on the application, one or the other version may be more suitable. We make use of both versions in our analyses: when simulating the victim, we use the near-subset, because we want to have high confidence that the transactions we use for simulation are indeed CoinJoins. When simulating the adversary, we use the near-superset version because the cluster intersection attack is more robust to false positives than false negatives.

\subsection{Additional Tables}
\label{app:tables}

\begin{table}[H]
\centering
\begin{tabular}{$L{5cm}}
\hline
\rowstyle{\bfseries} Third party domain \\ \hlineB{6}
monkeykingcode.com	\\ 
chatid.com 			\\ 
blockchain.info 	\\ 
google.com 			\\ 
revjet.com 			\\ 
qrserver.com 		\\ 
bootstrapcdn.com 	\\ 
americanexpress.com \\ 
schutzklick.de		\\ 
chart.googleapis.com\\ 
nosto.com			\\ 
gopollen.com		\\ 
exponea.com			\\ \hline
\end{tabular}
\caption{List of third parties that receive transaction-relevant information despite the use of tracking protection}
\label{fig:third-parties-despite-tp}
\end{table}

\begin{table}[H]
\centering
\begin{tabular}{$L{3cm}^L{2cm}^L{2cm}}
\hline
\rowstyle{\bfseries} Third-party domain & w/o protection & w/ protection \\ \hlineB{6}
google-analytics.com & 27 & 0  \\ \hline
facebook.com      & 16 & 0 \\ \hline
doubleclick.net        & 8 & 0  \\ \hline
google.com & 8 & 8 \\ \hline
chart.googleapis.com & 8 & 8 \\ \hline
segment.io & 3 & 0\\ \hline
steelhousemedia.com & 3 & 0 \\ \hline
criteo.com & 3 & 0 \\ \hline
blockchain.info & 3 & 3 \\ \hline
hits.io & 2 & 0 \\ \hline
monetate.net  & 2 & 0  \\ \hline
\end{tabular}
\caption{Prevalence of third parties receiving transaction-relevant information on at least two websites, with or without tracking protection}
\label{fig:tx-relevant-third-party-prevalence}
\end{table}

\begin{table}[H]
\centering
\begin{tabular}{$L{2cm}^L{2cm}^L{2cm}}
\hline
\rowstyle{\bfseries} PII type & w/o protection & w/ protection \\ \hlineB{6}
username      & 76 & 39 \\ \hline
email        & 63  & 30 \\ \hline
firstname & 41 & 19 \\ \hline
lastname      & 29 & 17 \\ \hline
address & 15 & 9 \\ \hline
phone       & 13 & 7  \\ \hline
name    & 7 & 4 \\ \hline
company & 5 & 3 \\ \hlineB{4}
\rowstyle{\bfseries} Third parties receiving PII & 137 & 70 \\ \hline
\end{tabular}
\caption{Number of third parties receiving each type of PII, with or without tracking protection}
\label{fig:third-party-pii-counts}
\end{table}

\begin{table}[H]
\centering
\begin{tabular}{$L{2cm}^L{2cm}^L{2cm}}
\hline
\rowstyle{\bfseries} Merchant & Receipt page third parties &  Receipt page TPs w/ tx-relevant info \\ \hlineB{6}
adafruit.com      & 13 & 5 \\ \hline
baronfig.com      & 35 & 7 \\ \hline
digitalrev.com    & 20 & 9 \\ \hline
fancy.com        & 0  & 0 \\ \hline
giftoff.com     & 10 & 1 \\ \hline
givemethedirt.com & 33 & 5 \\ \hline
healthmonthly.co.uk & 19 & 1 \\ \hline
jenshansen.com     & 59 & 7 \\ \hline
newegg.com      & 46 & 22 \\ \hline
opendime.com    &  5  & 5 \\ \hline
overstock.com & 42 & 42 \\ \hline
petspyjamas.com & 36 & 11 \\ \hline
pi-supply.com    & 0 & 0  \\ \hline
readytogosurvival.com   & 32 & 3 \\ \hline
reddit.com          & 1 & 1 \\ \hline
reeds.com      & 0 & 0 \\ \hline
somethinggeeky.com & 0 & 0 \\ \hline
thepihut.com & 44 & 8 \\ \hline 
thisisground.com & 50 & 4 \\ \hline
tightstore.com & 32 & 4 \\ \hlineB{4}
\rowstyle{\bfseries} Third parties on receipt pages & 245 & 88 \\ \hline
\end{tabular}
\caption{Number of third parties on each merchant's payment receipt page, and the number of those third parties that also received transaction-relevant information}
\label{fig:confirmation-page-tps}
\end{table}

\subsection{Additional Figures}
\label{app:figures}

\begin{figure}[h]
\centering
\includegraphics[width=0.7\columnwidth]{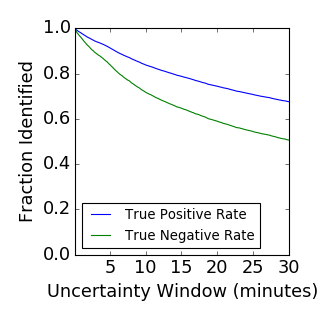}
\caption{Effect of payment time uncertainty on success rate when prices are sampled from the blockchain instead of from merchant websites. Compare to Figure \ref{fig:payment_time_uncertainty}.}
\label{fig:payment_time_uncertainty_bitcoin}
\end{figure}

\begin{figure}[h]
\centering
\includegraphics[width=0.7\columnwidth]{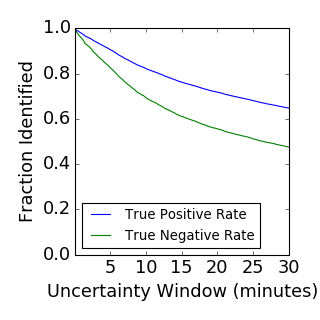}
\caption{Effect of payment time uncertainty on success rate when prices are doubled. Compare to Figure \ref{fig:payment_time_uncertainty}.}
\label{fig:payment_time_uncertainty_doubled}
\end{figure}

\begin{figure}[h]
\centering
\includegraphics[width=0.7\columnwidth]{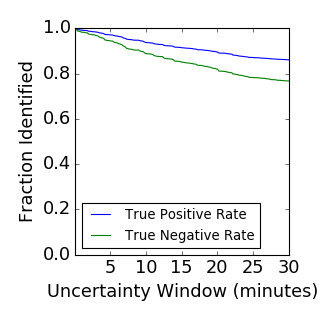}
\caption{Effect of payment time uncertainty on success rate when performed on Litecoin. Compare to Figure \ref{fig:payment_time_uncertainty}.}
\label{fig:payment_time_uncertainty_litecoin}
\end{figure}

\begin{figure}[h]
\centering
\includegraphics[width=0.7\columnwidth]{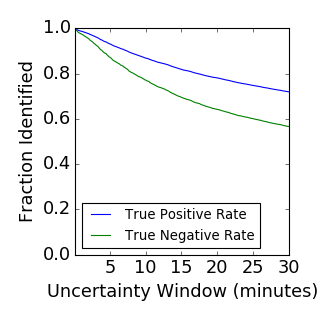}
\caption{Effect of payment time uncertainty on success rate against Bitpay transactions. Compare to Figure \ref{fig:payment_time_uncertainty}.}
\label{fig:payment_time_uncertainty_bitpay}
\end{figure}



\end{document}